\documentclass[reprint,twocolumn,showpacs,preprintnumbers,amsmath, aps,pre,amssymb]{revtex4-1}

\usepackage[algoruled]{algorithm2e}

\usepackage{times}
\usepackage{subcaption}
\usepackage{balance}
\usepackage{caption}
\usepackage{array}
\usepackage{multirow}
\usepackage{graphicx}
\usepackage{url}

\newcommand{\tg}[1]{}

\usepackage{expl3}
\ExplSyntaxOn
\newcommand\latinabbrev[1]{
  \peek_meaning:NTF . {
    #1\@}%
  { \peek_catcode:NTF a {
      #1.\@ }%
    {#1.\@}}}
\ExplSyntaxOff

\newcommand\eg{\emph{e.g.}}
\newcommand\ie{\emph{i.e.}}

\newcommand\bcmdtab{\noindent\bgroup\tabcolsep=0pt%
  \begin{tabular}{@{}p{10pc}@{}p{20pc}@{}}}
\newcommand\ecmdtab{\end{tabular}\egroup}

 \newcommand{\remove}[1]{}
\usepackage{color}

\newcommand{\nop}[1]{}

\usepackage{color}

\usepackage[nolist]{acronym}

\begin{document}

\label{firstpage}

\title{Web Content Extraction -- a Meta-Analysis of its Past and Thoughts on its Future}
\author{Tim Weninger}
 \email{tweninge@nd.edu}
\author{Rodrigo Palacios}
 \email{rodpl91@gmail.com}
\author{Valter Crescenzi}
 \email{crescenz@dia.uniroma3.it}
\author{Thomas Gottron}
 \email{gottron@uni-koblenz.de}
\author{Paolo Merialdo}
 \email{merialdo@dia.uniroma3.it} 
\affiliation{%
$^{\ast\dag}$Department of Computer Science and Engineering,
University of Notre Dame, Notre Dame, Indiana, USA\\
$^{\S}$Institute for Web Science and Technologies,
University of Koblenz-Landau, German\\
$^{\P}$Universit\`a Roma Tre,
Dipartimento di Ingegneria, Rome, Italy\\
}%
\date{\today}

\begin{abstract}
In this paper, we present a meta-analysis of several Web content extraction algorithms, and make recommendations for the future of content extraction on the Web. First, we find that nearly all Web content extractors do not consider a very large, and growing, portion of modern Web pages. Second, it is well understood that wrapper induction extractors tend to break as the Web changes; heuristic/feature engineering extractors were thought to be immune to a Web site's evolution, but we find that this is not the case: heuristic content extractor performance also tends to degrade over time due to the evolution of Web site forms and practices. We conclude with recommendations for future work that address these and other findings.
\end{abstract}

\maketitle

\tg{General Remark: Some of the figures and tables are out of place and do not appear close to the point in the paper where they are actually referenced.}

\section{Introduction}
The field of content extraction, within the larger pervue of data mining and information retrieval, is primarily concerned with the identification of the main text of a document, such as a Web page or Web site. The principle argument is that tools that make use of Web page data, \eg, search engines, mobile devices, various analytical tools, demonstrate poor performance due to noise introduced by text not-related to the main content~\cite{Finn2001,J:FI:2012:MartinG}.

In response the field of content extraction has developed methods that extract the main content from a given Web page or set of Web pages, \ie, a Web site~\cite{Liu2003,Zhai2005}. Frequently, these content extraction methods are based on pattern mining and the construction of well-crafted rules. In other cases, content extractors learn the general skeleton of a Web page by examining multiple Web pages in a Web site~\cite{Kushmerick1999,Bar-Yossef2002,Crescenzi2001,Crescenzi2008}. These two classes of content extractors are referred to as {\em heuristic} and {\em wrapper induction} respectively; and each class of algorithms have their own merits and disadvantages. Generally speaking, wrapper induction methods are more accurate than heuristic approaches, but require some amount of training data in order to initially induce an appropriate wrapper. Conversely, heuristic approaches are able to function without an induction step, but are generally less accurate. 

The main criticism of content extraction via wrapper induction is that the learned rules are often brittle and are unable to cope with even minor changes to a Web pages' template~\cite{Gibson2005}. When a Web site modifies its template, as they often do, the learned wrappers need to be refreshed by re-computing the expensive induction step. Certain improvements in wrapper induction attempt to induce extraction rules that are more robust to minor changes~\cite{Davulcu2000,Dalvi2009,ParameswaranDGR11}, but the more robust rules only delay the inevitable~\cite{Chidlovskii2006}. 
\tg{Do we have a citation for this claim? }

Heuristic approaches are often criticised for their lack of generality. That is, heuristics that may work on a certain type of Web site, say a news agency, are often ill suited for business Web sites or message boards, etc. Most approaches also ignore the vast majority of the Web pages that dynamically download or incorporate content via external reference calls during the rendering process, \eg, CSS, JavaScript, images.

\tg{The beginning of the next paragraph sounds rather defensive and provides a basis for criticism. I do not think this statement is needed, as before we already say, that CE approaches can be divided into two classes \emph{in general}.}
The goal of this paper is not to survey the whole of content extraction, so we resist the temptation to verbosely compare and contrast the numerous published methods. Rather, in this paper we make a frank assessment on the state of the field, provide an analysis of content extraction effectiveness over time, and make recommendations for the future of content extraction.

In this paper we make three main contributions:

\begin{enumerate}
\item We define the vectors of change in the function and presentation of content on the Web, \tg{Not clear to me: what is meant with vector of change}
\item We examine the state of content extraction with respect to the ever changing Web, and
\item We perform a temporal evaluation on various content extractors
\end{enumerate}

Finally, we call for a change in the direction of content extraction research and development. 

The evolution of Web practices is the central to the theme of this paper. A scientific discipline ought to strive to have some invariance in the results over time. Of course, as technology changes, our study of it must also change as well. With this in mind, one way to determine the success of a model is to measure its stability or durability as the input changes over time.

To that end, we present the results of a case study that compares content extraction algorithms, both old and new, on an evolving dataset. The goal is to identify which measures, if any, are invariant to the evolution of Web practices. 

\begin{table}[h]
\centering
\begin{tabular}{l}
 Web site \\ \hline
 \url{news.bbc.co.uk}\\
 \url{cnn.com}\\
 \url{news.yahoo.com}\\
 \url{thenation.com}\\
 \url{latimes.com}\\
 \url{entertainment.msn.com}\\
 \url{foxnews.com}\\
 \url{forbes.com}\\
 \url{nymag.com}\\
 \url{esquire.com}\\
\end{tabular}
\caption{Dataset used in case study. 25 Web pages crawled from each Web site per lustrum (5-year period), over 4 lustra and 10 Web sites totals 1,000 Web pages.}
\label{tab:dataset}
\end{table}

To that end, we collected a dataset of 1000 Web pages from 10 different domains, listed in Table~\ref{tab:dataset}, where each domain has a set of pages from years 2000, 2005, 2010, and 2015. There are 25 HTML documents per lustrum (\ie., 5-year period), for a total of 100 documents per Web site. The documents were automatically and manually gathered from two types of sources: archives\footnote{WayBack Machine - http://archive.org/web/} and the original websites themselves for the 2015 lustrum.

We review the evolution that has occurred in Web content delivery and extraction, referring explicitly to recent changes that undermine the effectiveness of exiting content extractors. To show this explicitly we perform a large case study wherein we compare the performance over time of several content extraction algorithms. Based on our findings we call for a change in content extraction research and make recommendations for future work.

\tg{any remark on how these contributions are made? Methodology? Might be of interest to the reader at this point.}

\begin{figure*}[!htbp]
    \centering
    \begin{minipage}[t]{0.31\textwidth}
        \includegraphics[width=\textwidth]{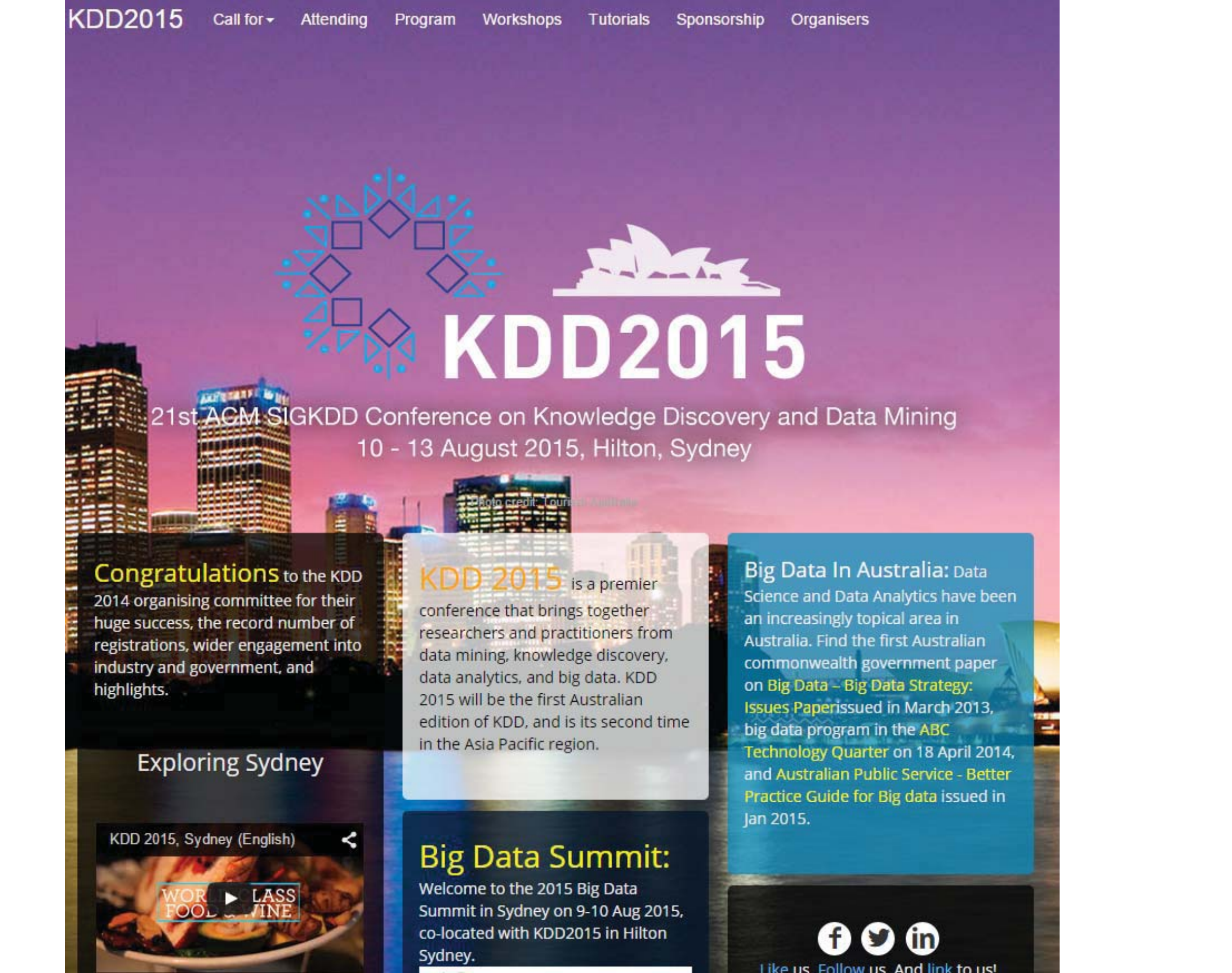}
    \end{minipage}%
    \begin{minipage}[t]{0.31\textwidth}
        \includegraphics[width=\textwidth]{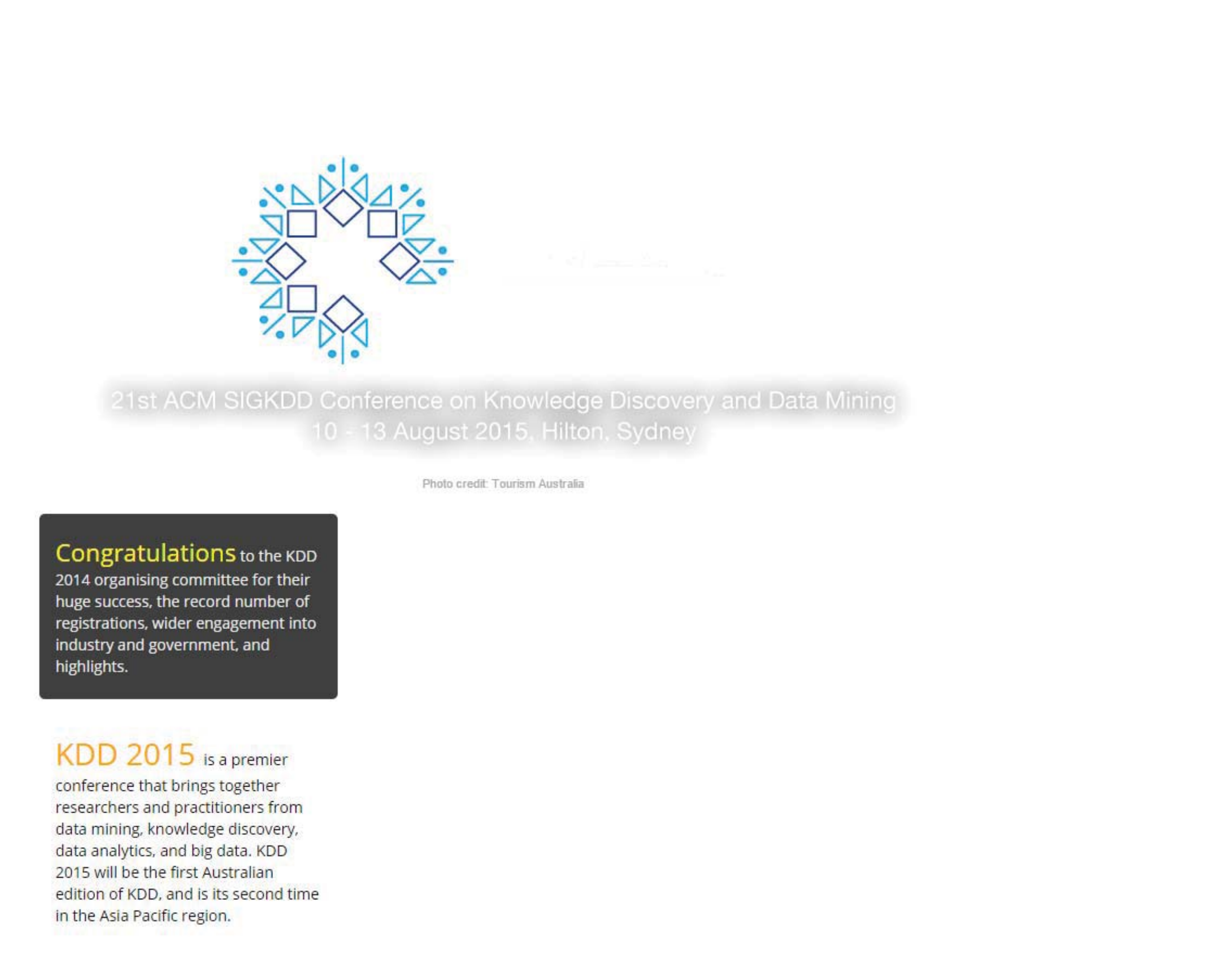}
    \end{minipage}
    \begin{minipage}[t]{0.31\textwidth}
        \includegraphics[width=\textwidth]{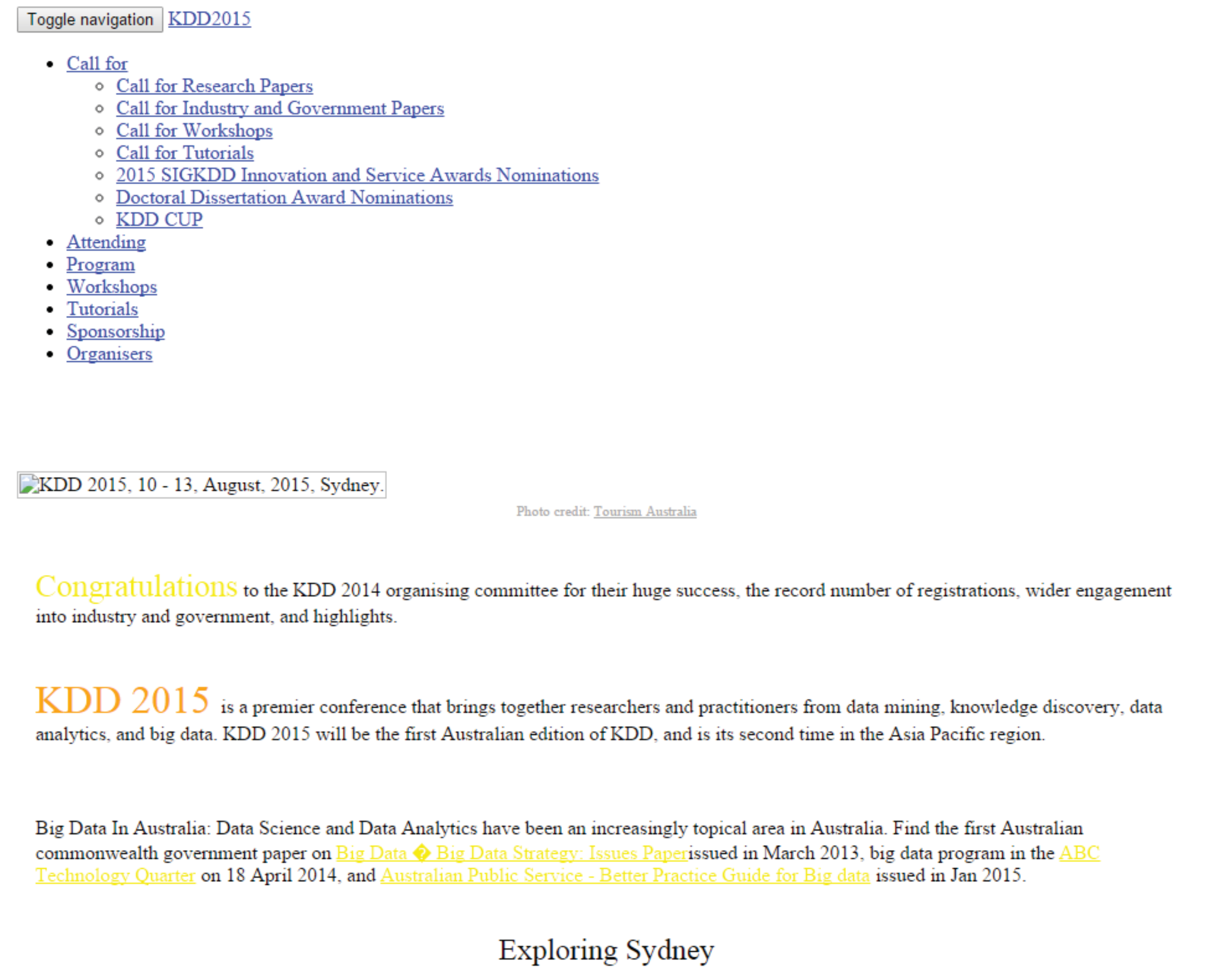}
    \end{minipage}
    \caption{The Web page of \protect\url{http://www.kdd.org/kdd2015/} fully rendered in a modern Web browser (Left). Web page with JavaScript disabled (Middle). Downloaded Web page HTML, statically rendered without any external content (Right). Most extractors operate on the Web page on the right.}
    \label{fig:kddwebpage}
\end{figure*}

\begin{table*}[t]
\centering
\begin{tabular}{r|r r r r r r r r}
        &  \texttt{src}     & \texttt{link}   & \texttt{iframe}    & \texttt{script}  & js      & jquery  & css     & size \\ \hline
 2000   &  37.092  & 1.152  & 0.388	    & 7.600	 & 6.588	& 0.908	  & 1.828   & 39,121.90 \\
 2005   &  61.812  & 2.528	& 0.408	    & 21.200 & 14.280	& 0.944	  & 2.612   & 52,633.82\\
 2010   &  57.976  & 10.104	& 1.408 	& 44.044 &	24.096	& 1.000	  & 10.468  & 81,033.89\\
 2015   &  49.396  & 18.256	& 10.032	& 40.652 &	37.052	& 0.620	  & 11.692  & 174,801.64 \\
\end{tabular}
\caption{The mean-average occurrences of certain HTML tags and attributes that represent ancillary source files in our dataset of 1,000 news Web pages over 4 equal sized lustra (5-year periods). The use of external content and client-side scripting has been growing quickly and steadily.}
\label{tab:tags}
\end{table*}


\section{Evolving Web Practices}
\label{sec:evolve}
\tg{This section references repeatedly to the dataset without having introduced or described it. This is inconsistent. Also, there are some hints on results of the experiments, which creates confusion and makes the paper seem not well structured.}

We begin with the observation that the content delivery on the Web has changed dramatically since it was first conceived. The case for content extraction is centered around the philosophy that HTML is a markup language that describes how a Web page ought to {\em look}, rather than what a Web page contains. Here, the classic form versus function debate is manifest. Yet, in recent years the Web has seen a simultaneous marriage and divorce of form and function with the massive adoption of scripting languages like JavaScript and with the finalization of HTML5.

In this section we argue that because Web technologies have changed, the way we perform and evaluate content extraction must also change.

\subsection{Evolution of Form and Function}

\vspace{2mm}\textbf{JavaScript.} Nearly all content extraction algorithms operate by downloading the HTML of the Web page(s) under consideration, and only the HTML. In many cases, Web pages refer directly or indirectly to dozens of client side scripts, \ie, JavaScript files, that may be executed at load-time. Most of the time content extractors do not even bother to download referenced scripts even though JavaScript functions can (and frequently do) completely modify the DOM and content of the downloaded HTML. Indeed, most of the spam and advertisements that content extraction technologies explicitly claim to catch are loaded via JavaScript and are therefore not part of most content extraction testbeds. 

\vspace{2mm}\textbf{CSS.} Style sheets pose a problem similar in nature to JavaScript in that structural changes to the displayed content on a Web site are frequently performed by instructions embedded in cascading style sheets. Although CSS instructions are not as expressive as JavaScript functions -- they were built for different purposes -- the omission of a style sheet often severely affects the rendering of a Web page.

Furthermore, many of the content extractors described earlier rely on formatting hints that live within HTML in order to perform effective extraction. Unfortunately, the ubiquitous use of CSS removes many of the HTML hints that extractors depend upon. Using CSS, it is certainly possible that a complex Web site is made entirely of \texttt{div}-tags. 

\vspace{2mm}\textbf{HTML5.} The new markup standards introduced by HTML5 include many new tags, including \texttt{main}, \texttt{article}, \texttt{header}, etc., meant to specify the semantic meaning of content. Widespread adoption of HTML5 is in progress, so it is unclear whether and how the new markup languages will be used or what the negative side effects will be, if any.

\tg{Here I would strongly object! HTML has always been a document structure language -- not a layout language! It has been often abused for layout purposes, and HTML4 already has declared certain purely layout oriented elements as deprecated}

The semantic tags in HTML5 are actually a severe departure form the original intent of HTML. That is, HTML4 was originally meant to be a markup for the structure of the Web page, not a description language. Indeed the general lack of semantic tags is one of the main reasons why content extraction algorithms were created in the first place.

\tg{The latter is partially correct: There are some shallow and low level semantics in HTML4. However, nothing comparable to the novel elements in HTML5}

Further addition of semantics into HTML markup is provided by the \url{schema.org} project. Schema.org is a collaboration among the major Web search providers to provide a unified description language that can be embedded into HTML4/5 tag attributes. Web site developers can use these tags to encode what HTML data represents, for example, a Person-\texttt{itemtype}, which may have a name-\texttt{itemprop}, can then be used by search engines and other Web-services to built intelligent analytics tools. Other efforts to encode semantic meaning in HTML can be found in the \url{Microformats.org} project, the Resource Description Framework in Attributes (RDFa) extension to HTML5, and others.

\begin{table}[h]
\centering
\begin{tabular}{r || r r r | r r}
     & \texttt{itmscp} & \texttt{itmtp} & \texttt{itmprp} & \texttt{sctn} & \texttt{artcl}  \\ \hline
Mean & 162.2 & 157.8 & 899.0 & 261.0 & 403.4 \\
Median & 65.5 &	54.5 & 374.5 & 25 & 166.5 \\
\end{tabular}
\caption{Mean and Median number of occurrences of semantic tags from \protect\url{schema.org}: \texttt{itemscope}, \texttt{itemtype} and \texttt{itemprop} tags, and from HTML5: \texttt{article} and \texttt{section} found in 2015-subset of the dataset. Semantic tags are only found in dataset from 2015.}
\label{tab:semantic}
\end{table}

Table~\ref{tab:semantic} shows the mean and median number of Schema.org and HTML5 semantic tags in our 2015 dataset. We find that 9 out of 10 Web sites we crawled had adopted the Schema.org tagging system, and that 9 out of 10 Web sites had adopted the \texttt{section} and \texttt{article} tags from HTML5 (8/10 adopted both Schema.org and HTML5).

The advent and widespread adoption of HTML5 and Schema.org decreases the need for many extraction tools because the content or data is explicitly marked and described in HTML.

\vspace{2mm}\textbf{AJAX.} Often, modern Web pages are delivered to the client without the content at all. Instead, the content is delivered in a separate JSON or XML message via AJAX. These are not rare cases, as of April 2015, Web Technologies research finds that AJAX is used within 67\% of all Web sites\footnote{\protect\url{ http://w3techs.com/technologies/overview/javascript_library/all}. Accessed May 6, 2015.}. Thus, it is conceivable that the vast majority of content extractors over estimate their effectiveness in 67\% of the cases, because a large portion of the final, visually-rendered Web page is not actually present in the HTML file. \tg{At this point in the document it is not clear for the reader, why this causes an overestimation of performance...}

In fact, in our experiments we find that the most frequent last-word found by many content extractors on NY Times articles is ``loading...''

Table~\ref{tab:tags} shows the mean-average number of occurrences of certain HTML tags and attributes that represent ancillary source files in our dataset of 1,000 Web pages. In this table, \texttt{src} refers to the occurrence of the common tag attribute which can refer to a wide range of file types. \texttt{link} refers to the occurrence of the \texttt{<link>} HTML tag which frequently (although not necessarily) references external CSS files. \texttt{iframe} refers to the occurrence of the HTML tag which is used to embed another HTML document into the current HTML document. \texttt{script} refers to the occurrence of the HTML tag which is used to denote a client-side script such as (but not necessarily) JavaScript. \texttt{js} refers to the occurrences of externally referenced JavaScript files; \texttt{css} similarly refers to the occurrences of externally referenced CSS files. The \texttt{jquery} column shows the percentage of Web pages that employ AJAX via the jQuery library; alternative AJAX libraries were found but their occurrence rates were very small.

In many ways the above observations show that the Web is trending towards a further decoupling of form from content: JavaScript decouples the rendered DOM from the downloaded HTML, CSS similarly separates the final presentation from the downloaded HTML, and AJAX allows for the HTML and extractable content to be separate files entirely. Yet, despite these trends, most content extraction methodologies rely on extractions from statically downloaded HTML files. \tg{What is a "statically downloaded" HTML file? The term does not make sense. Do you mean a static HTML file?}

An example of why this should be considered a bad practice is highlighted in Figure~\ref{fig:kddwebpage} where the Web page \url{http://kdd.org/kdd2015} is shown rendered in a browser (at left), rendered without JavaScript (center), and rendered with only the static HTML document (at right). The information conveyed to the end user is presented in its complete form in the rendered version; thus, content extractors should strive to operate within the fully rendered document (at left), instead of the HTML-only extraction as is the current practice (at right).

\tg{I like the example, but I am not sure if the point to be made is communicated clearly enough. Would it make sense to rather talk about the information available to a human end user (via visual representation) vs. the purely text and structure based information of the raw and static HTML file?}

\subsection{Keeping Pace with the Changing Web}

Web presentation has evolved in remarkable ways in a very short time period. Content Extraction algorithms have attempted to keep pace with evolving Web practices, but many content extraction algorithms quickly become obsolete.

Counter-intuitively, it seems that as although the number of Web sites has increased, the variety of presentation styles has actually decreased. For a variety of reasons, most Web pages within the same Web site look strikingly similar. Marketing and brand-management often dictate that a Web site maintains style distinct from competitors, but are similar to other pages in the same Web site. 

\tg{Why are the following paragraphs part of the section "Evolution of Form"? They do not discuss the evolution of the Web at all...}

\vspace{2mm}\textbf{Wrapper Induction.} The self-similarity of pages in a Web site stem from the fact that the vast majority of Web sites use scripts to generate Web page content retrieved from backend databases. Because of the structural similarity of Web pages within the same Web site, it is possible to reverse engineer the page generation process to find and remove the Web site's skeleton, leaving only the content remaining~\cite{Kushmerick1999,Bar-Yossef2002,Crescenzi2001,Crescenzi2008}.

A wrapper is induced on one Web site at a time and typically needs only a handful of labelled examples. Once trained the learned wrapper can extract information at near-perfect levels of accuracy. Unfortunately, the wrapper induction techniques assume that the Web site template does not change. Even the smallest of tweaks to a Web site's template or the database schema breaks the induced wrapper and requires retraining. Attempts to learn robust wrappers, which are immune to minor changes in the Web page template have been somewhat successful, but even the most robust wrapper rules eventually break~\cite{Dalvi2009,Gibson2005}.

\vspace{2mm}\textbf{Heuristics and Feature Engineering.} Rather than learning rigid rules for content extraction, other works have focused on identifying certain heuristics as a signal for content extraction. The variety of the different heuristics is impressive, and the statistical models learned through a combination of various features may, in many cases, perform comparable to extractors based on wrapper induction. 

Each methodology and algorithm was invented at a different time in the evolution of the Web and looked at different aspects of the Web content. From the myriad of options we selected 11 algorithms from different time periods. They are listed in Table~\ref{tab:cealgs}. 

\begin{table}[h]
\centering
\begin{tabular}{r|l|c}
Algorithm & & Year  \\ \hline
Body Text Extractor (BTE) & \cite{Finn2001} & 2001 \\
Largest Size Increase (LSI) & \cite{Han2001} & 2001 \\
Document Slope Curve (DSC) & \cite{Pinto2002} & 2002 \\
Link Quota Filter & \cite{Mantratzis2005} & 2005 \\
K-Feature Extractor (KFE) & \cite{Debnath2005} & 2005 \\
Advanced DSC (ADSC) & \cite{Gottron2008} & 2007 \\
Content Code Blurring (CCB) & \cite{Gottron2008a} & 2008 \\
RoadRunner$^{\ast}$ (RR) & \cite{Crescenzi2008} & 2008 \\
Content Extraction via Tag Ratios (CETR) & \cite{Weninger2010a} & 2010 \\
BoilerPipe & \cite{Kohlschutter2010} & 2010 \\
Eatiht & \cite{Palacios2015} & 2015 \\
\end{tabular}
\caption{Content extraction algorithms, with their citation and publication date. $^{\ast}$RoadRunner is a wrapper induction algorithm; all others are heuristic methods.}
\label{tab:cealgs}
\end{table}

Each algorithm, heuristic, model or methodology is predicated on the form and function of the Web at the time of its development. Each was evaluated similarly on the state of the Web that existed at the time, presumably, just before publication. Furthermore, each algorithm does not consider JavaScript, CSS, or AJAX changes to the Web page, therefore the majority of the Web page may not actually be present for extraction, as is the case in Figure~\ref{fig:kddwebpage}.

\begin{figure*}[t]
\centering
\begin{subfigure}[t]{0.45\textwidth}
    \centering
    \includegraphics[width=\textwidth]{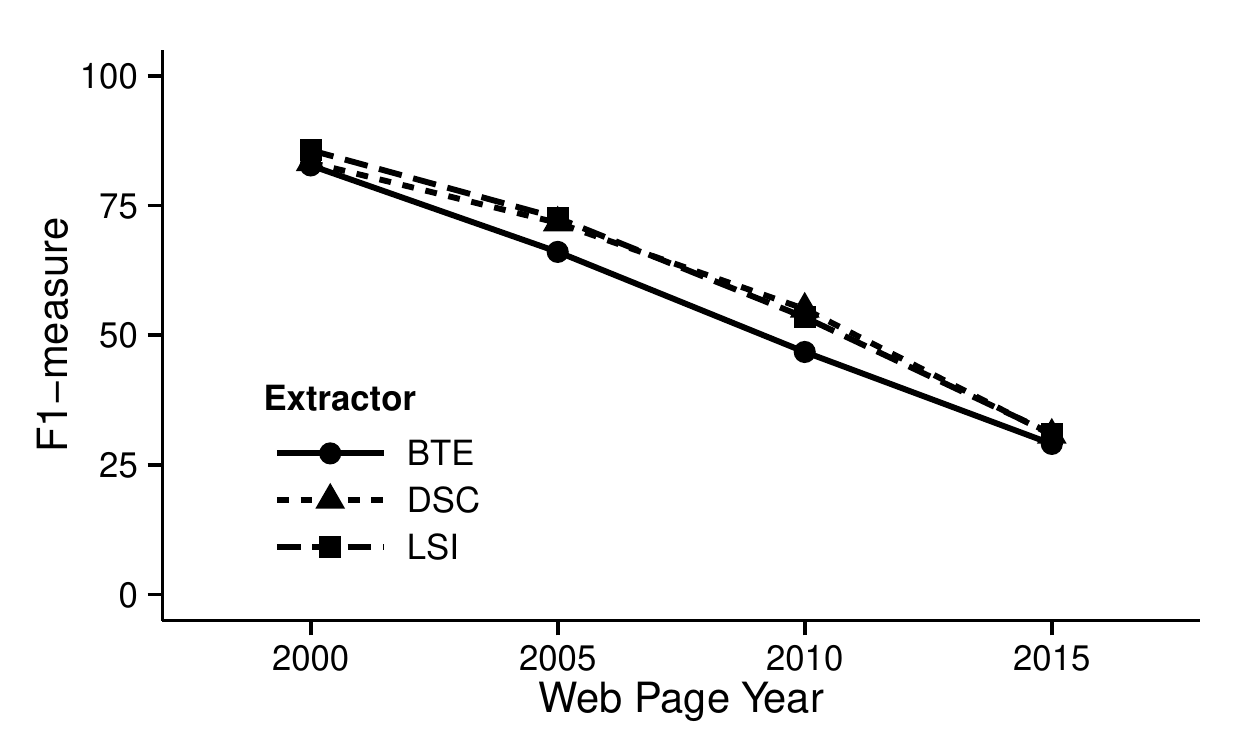} 
    \caption{\emph{ca}. 2000}
    \label{fig:a}
\end{subfigure}
\begin{subfigure}[t]{0.45\textwidth}
    \centering
    \includegraphics[width=\textwidth]{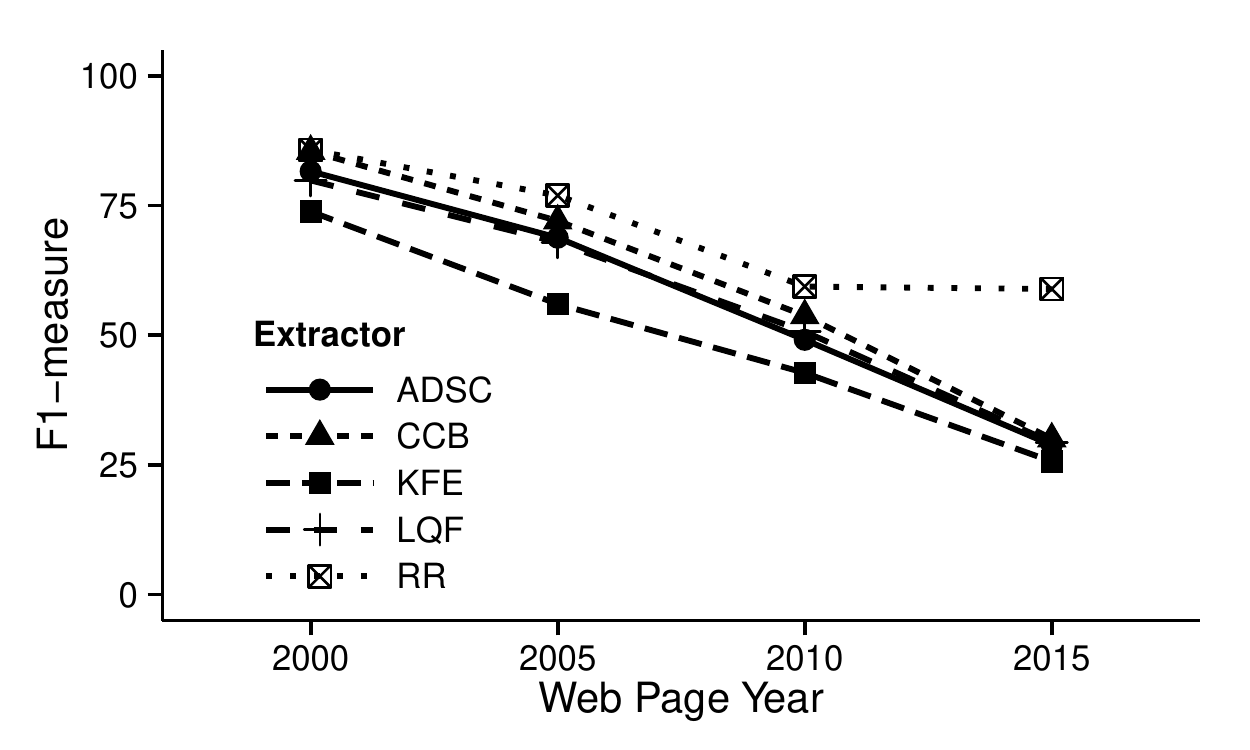} 
    \caption{\emph{ca}. 2005}
    \label{fig:b}
\end{subfigure}
\begin{subfigure}[t]{0.45\textwidth}
    \centering
    \includegraphics[width=\textwidth]{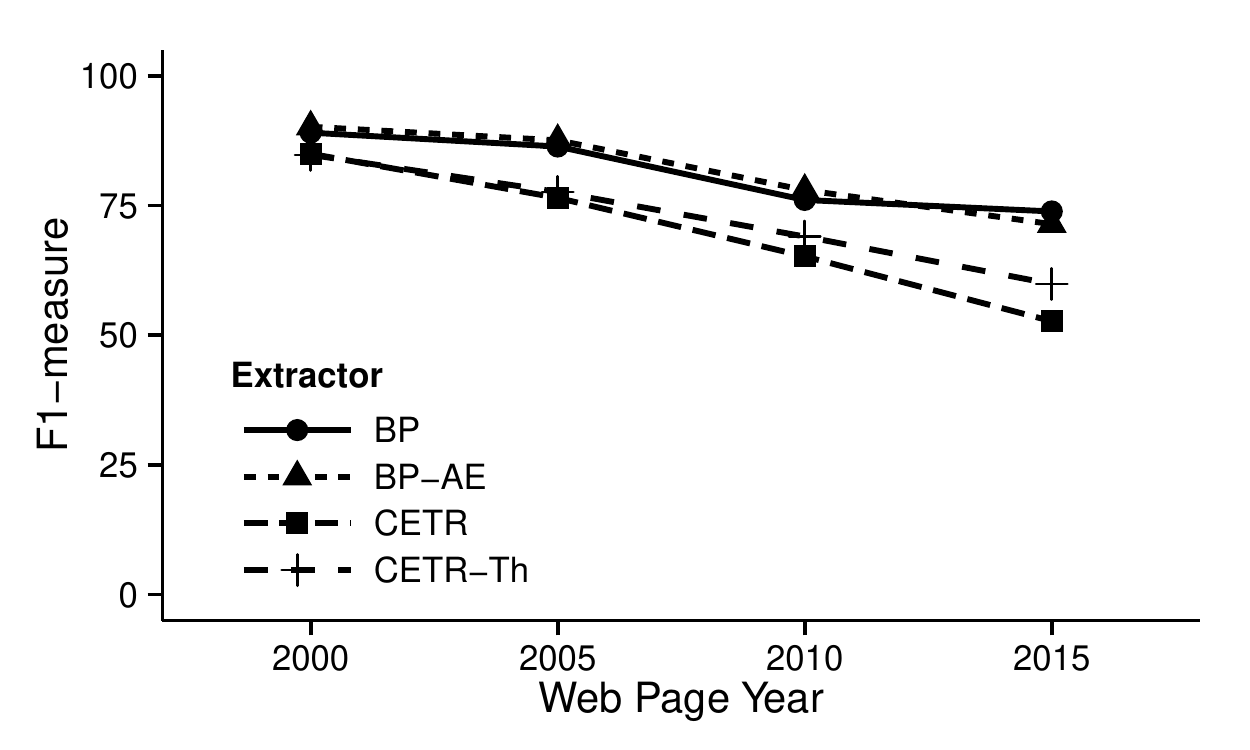} 
    \caption{\emph{ca}. 2010}
    \label{fig:c}
\end{subfigure}
\begin{subfigure}[t]{0.45\textwidth}
    \centering
    \includegraphics[width=\textwidth]{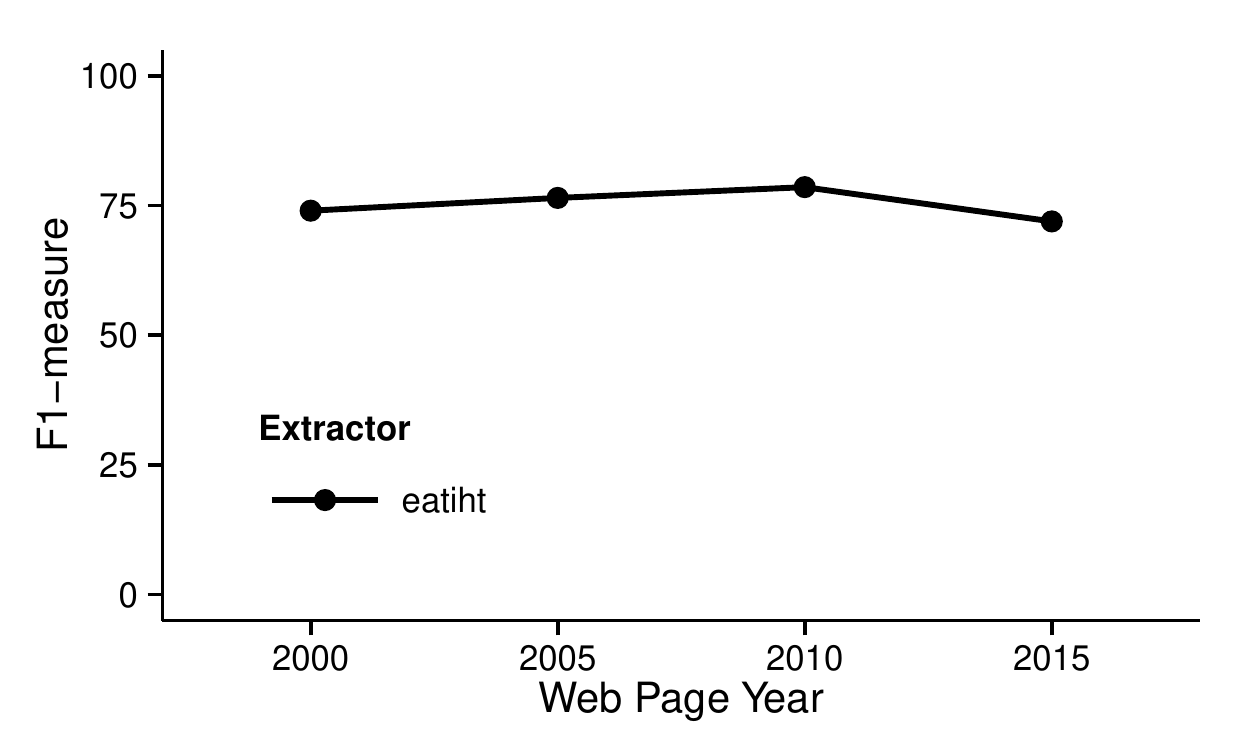} 
    \caption{\emph{ca}. 2015}
    \label{fig:d}
\end{subfigure}
    \caption{F$_1$-measure for various extractor cohorts by lustrum (5-year period).} 
\end{figure*}

\section{Case Study}
\label{sec:method}

We present the results of a case study that compares content extraction algorithms, both old and new, on an evolving dataset. The goal is to test the performance variability of content extractors over time as Web sites evolve. So, for each Web page of each lustrum of each Web site, a gold-standard dataset was created manually by the second author. Each Web content extractor attempted to extract the main content from the Web page.

For the first seven content extractors in Table~\ref{tab:cealgs}, we used the implementation from the CombineE System~\cite{Gottron2008}. The Eatiht, BoilerPipe and CETR implementations are all available online. BoilerPipe provides a standard implementation as well as an article extractor (AE), Sentence extractor (Sen), an extractor trained on data from KrdWrd-Canola corpus\footnote{\protect\url{https://krdwrd.org/trac/raw-attachment/wiki/Corpora/Canola/CANOLA.pdf}}, and two ``number of words'' extractors: a decision tree induced extractor (W) and a decision tree induced extractor manually tuned to at least 15 words per content area (15W). CETR has a default algorithm as well as a threshold option based on the standard deviation of the tag ratios (Th), and a 1 dimension clustering option (1D). See the respective papers for details.

An attempt was made to induce wrappers using the Roadrunner wrapper induction system~\cite{Crescenzi2008}, which was successful on each set of 25 Web pages, but performed very poorly on the proceeding lustrum. Wrapper-breakage is a well known problem for wrapper induction techniques~\cite{Dalvi2009,Gibson2005}. A five-year window is too long for any wrapper to continue to be effective. Thus Roadrunner had to be trained and evaluated slightly differently. In this case we manually identified Web pages that have very similar HTML structure and learned a wrapper on those few pages. In most cases 90-95\% of the Web pages in a single domain could be used to generate a wrapper, but in 2 Web sites only about half of the Web pages were found to have the same style and were useful for training. We used the induced wrapper to extract the content from the Web pages on which it was trained. 

We emphasize that our methodology follows that of most content extraction methodologies. Namely, we download the raw HTML of the Web page and perform content extraction on \emph{only} that static HTML. We further emphasize that this ignores a very large portion of the overall rendered Web page -- renderings that are increasingly reliant on external sources for content and form via AJAX, stylesheets, iframes, etc. The disadvantages of this methodology are clear, but we are beholden to them because the existing extractors require only static HTML.

\subsubsection{Evaluation}

We employ standard content extraction metrics to compare the performance of different methods. Precision, recall and F$_1$-scores are calculated by comparing the results/output of each methods to a hand-labeled gold standard. The F$_1$-scores are computed as usual and all results are calculated by averaging each of the metrics over all examples.

The main criticism of these metrics is that they are likely to be inflated. This is because every word in a document is considered to be distinct even if two words are lexically the same. This makes it impossible to align words with the original page and therefore forces us to treat the hand labeled content and automatically extracted content as a bag of words, \eg{i.e.}, where two words are considered the same if they are lexically the same. The bag of words measurement is more lenient and as a result scores may be inflated.

\tg{There are alternative approaches to Bag of Words (BoW). I typically used the Word sequence (WS) and computed the overlap as the longest common subsequence. However, in one of my experiments I did not see big deviations between BoW and WS.}

The CleanEval competition has a hand-labeled gold standard as well from a shared list of 684 English Web pages and 653 Chinese Web pages downloaded in 2006 by ``[collecting] URLs returned by making queries to Google, which consisted of four words frequent in an individual language''\cite{Baroni2008}. CleanEval uses a different approach when computing extraction performance. Their scoring method is based on a word-at-a-time version of the Levenshtein distance between the extraction algorithm and the gold standard divided by the alignment length. 

\tg{Not a good argument -- Levenshtein is computationally as  expensive as longest common subsequence. And I ran such an evaluation 8 years ago and old machinery ...}

\begin{table*}[t]
\centering
\begin{tabular}{l l r | r ||r r r | r r r | r r r | r r r}
&&        &           & \multicolumn{12}{c}{Lustrum}\\
&&        &           & \multicolumn{3}{c|}{2000}& \multicolumn{3}{c|}{2005}& \multicolumn{3}{c|}{2010}& \multicolumn{3}{c}{2015} \\
&& Extractor & Year & Prec & Rec & Acc & Prec & Rec & Acc & Prec & Rec & Acc & Prec & Rec & Acc \\ \cline{2-16}
&& All Text	&	--	&	45.65	&	100	& 45.65	&	38.33	&	100	& 38.33	&	25.78	&	100	&	25.78	&	20.14	&	100	&	20.14\\ \cline{2-16}
\multirow{18}{*}{\rotatebox[origin=c]{90}{Cohort~~~~}}&\multirow{3}{*}{\rotatebox[origin=c]{90}{2000}}&BTE	&	2001	&	76.36	&	92.74	&	82.74	&	58.15	&	89.37	&	72.22	&	34.32	&	88.67	&	53.92	&	20.67	&	85.47	&	44.05\\
&&LSI	&	2001	&	83.37	&	89.79	&	87.87	&	64.19	&	88.71	&	77.32	&	43.35	&	80.47	&	65.7	&	23.66	&	77.51	&	48.56\\
&&DSC	&	2002	&	85.42	&	83.25	&	86.2	&	66.88	&	82.98	&	77.55	&	46.37	&	75.74	&	71.09	&	23.89	&	72.97	&	50.15\\ \cline{2-16}
&\multirow{5}{*}{\rotatebox[origin=c]{90}{2005}}&KFE	&	2005	&	74.21	&	75.88	&	83.34	&	50	&	69.95	&	70.71	&	35.28	&	63.45	&	65.4	&	19.78	&	64.92	&	47.72\\
&&LQF	&	2005	&	71.11	&	93.81	&	81.49	&	56.57	&	92.39	&	72.23	&	38.68	&	85.44	&	61.94	&	20.91	&	84.47	&	45.24\\
&&ADSC	&	2007	&	74.39	&	92.91	&	83.68	&	57.68	&	91.36	&	73.51	&	36.95	&	86.74	&	59.28	&	20.27	&	85.59	&	44.25 \\
&&CCB 	&	2008	&	85.28	&	86.95	&	88.26	&	65.79	&	84.91	&	77.78	&	44.45	&	77.06	&	68.95	&	22.92	&	74.43	&	48.72\\
&&RR	&	2008	&	81.97	&	92.11	&	88.96	&	70.73	&	89.75	&	86.54	&	61.32	&	76.57	&	70.82	&	47.75	&	86.25	&	79.70\\ \cline{2-16}
&\multirow{9}{*}{\rotatebox[origin=c]{90}{2010}}&CETR	&	2010	&	86.74	&	85.18	&	88.98	&	76.05	&	82.08	&	85.13	&	59.01	&	81.32	&	80.79	&	54.66	&	67.86	&	88.03\\
&&CETR-1D	&	2010	&	85.3	&	85.62	&	88.55	&	76.23	&	82.64	&	85.41	&	59.31	&	80.35	&	80.89	&	56.57	&	67.13	&	87.78\\
&&CETR-Th	&	2010	&	89.92	&	81.95	&	89.16	&	82.1	&	77.52	&	86.74	&	65.63	&	78.31	&	84.21	&	57.42	&	72.76	&	89.6\\
&&BP	&	2010	&	93.51	&	85.92	&	92.26	&	91.84	&	82.64	&	92.45	&	79.12	&	75.72	&	88.86	&	83.17	&	68.84	&	93.74\\
&&BP-AE	&	2010	&	94.76	&	87.11	&	92.86	&	92.97	&	84.25	&	92.54	&	94.99	&	69.39	&	91.35	&	85.79	&	63.21	&	92.96\\
&&BP-Sen	&	2010	&	97.37	&	84.43	&	92.78	&	97.47	&	81.71	&	93.53	&	97.19	&	66.84	&	91.26	&	89.06	&	61.33	&	93.09\\
&&BP-Canola	&	2010	&	93.43	&	87.36	&	92.58	&	88.09	&	84.56	&	90.97	&	77.33	&	77.47	&	88.71	&	68.74	&	71.47	&	92.02\\
&&BP-15W	&	2010	&	94.5	&	83.7	&	91.55	&	89.09	&	80.62	&	90.22	&	82.1	&	74.04	&	89.37	&	73.89	&	68.51	&	92.4\\
&&BP-W	&	2010	&	91.45	&	88.83	&	92.51	&	88.31	&	86.12	&	91.76	&	81.79	&	78.54	&	89.58	&	83.31	&	70.84	&	93.97\\ \cline{2-16}
&\multirow{2}{*}{\rotatebox[origin=c]{90}{2015}}
&eatiht	&	2015	&	81.89	&	76.3	&	80.17	&	82.04	&	80.04	&	83.76	&	93.75	&	69.18	&	91.13	&	88.48	&	62.93	&	94.39\\ 
&&&&&&&&&&&&&&&\\

\end{tabular}
\caption{Precision, recall and accuracy breakdown by lustrum (\ie, the 5-year period in which data was collected) and cohort (\ie, the set of extractors that were developed in the same time period)}
\label{tab:raw}
\end{table*}

\subsection{Results}

First, we begin with a straightforward analysis of the results of each algorithm on the dataset. Figure~\ref{fig:a}--\ref{fig:d} shows the F$_1$-measure for each lustrum, \ie, each 5-year time period, organized by extractor cohort. For example, the BTE-extractor was published in 2001, and is therefore part of the \emph{ca}. 2000 cohort of extractors; it's performance is illustrated in Figure~\ref{fig:a}. The eatiht-extractor was published in 2015 and is therefore part of the \emph{ca}. 2015 cohort of extractors, and is illustrated in Figure~\ref{fig:d}.

The shape of the performance curves in Figure~\ref{fig:a}--\ref{fig:d} over time exactly demonstrate the primary thesis of this paper: extractors quickly become obsolete. 

\begin{figure}[t]
\centering
    \includegraphics[width=.48\textwidth]{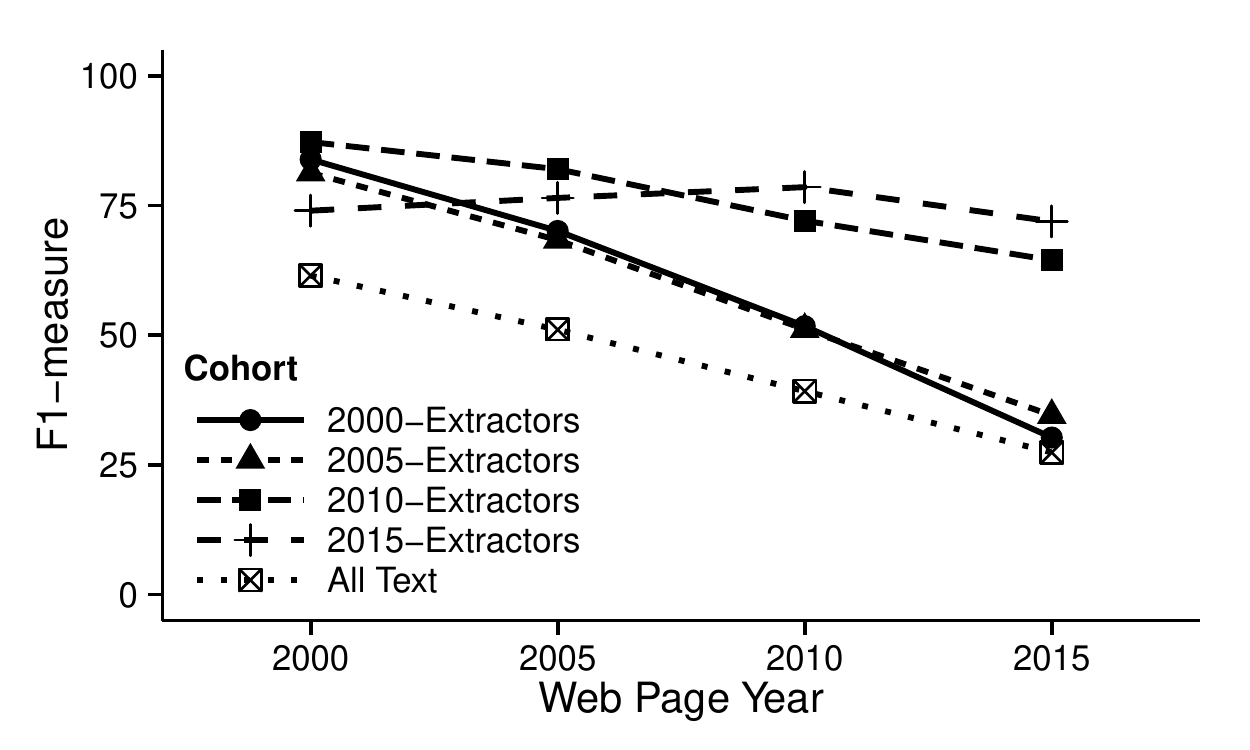} 
\caption{Mean average F$_1$ measure per cohort over each lustrum.}
\label{fig:ext}
\end{figure}

Indeed, Figure~\ref{fig:ext} averages the F$_1$-measure of each cohort and plots their aggregate performance together. We can clearly see that all of the extractor cohorts begin at approximately the same performance on Web page data from the year 2000, but the performance quickly falls as the form and function of the Web pages change. As a naive baseline, we also measure the results if all non-HTML text was extracted and treated as content; in this case, the F$_1$-measure is buoyed by the perfect recall score, but the precision and accuracies are bad as expected.

2015-extractors are most invariant to changes in the Web because the developers likely created the extractor knowing the state of the Web in 2015 and with an understanding of the history of the Web. 2010-extractors perform well on data from 2010 and prior, but were unable to adapt to unforeseen changes that appeared in 2015. Similarly extractors from 2005 performed well on data from 2005 and prior, but did not predict Web changes and quickly became obsolete. 

The F$_1$-measure is arguably the best single performance metric to analyze this type of data, however, individual precision, recall and accuracy considerations may be important to various applications. The raw scores are listed in Table~\ref{tab:raw}.

We find that extractors from 2000 and 2005 have a steep downward trend and extractors from 2010 also has a downward trend, although not as steep. Only the 2015 extractor performs steadily. These results indicate that changes Web design and implementation has adversely affected content extraction tools.

The semantic tags found in new Web standards like HTML5 may be one solution to the falling extractor performance. Table~\ref{tab:article} demonstrates surprisingly good extraction performance by extracting only (and all of) the text inside the \texttt{article} tag from the 2015 lustrum as the articles content. Compared to the results from the complex algorithms shown in Table~\ref{tab:raw} the simple HTML5 extraction rule shows reasonable results with very little effort.

\begin{table}[h]
\centering
\begin{tabular}{r | r r r}
        & Precision & Recall & Accuracy  \\ \hline
Mean     & 57.3 & 67.3 & 82.4 \\
Median   & 60.7 & 72.3 & 83.4 \\

\end{tabular}
\caption{Extraction results using only HTML5 \texttt{article} tags.}
\label{tab:article}
\end{table}

This further demonstrates that the nature of the Web is changing, and as a result, our thinking about content extraction must change too.

\subsection{Discussion}

The main critique of wrapper induction methods is that they frequently require re-training. In response many heuristic/feature engineering approaches have been developed that are said to not require training and simply work out of the box. 

These results underscore a robustness problem in Web content extraction. Ideally, Web science research should be at least partially invariant to change. If published content extractors are to be adopted and widely used they ought to be able to withstand changing Web standards. Wrapper induction techniques admit this problem; however, we find that heuristic content extractors are prone to obsolescence as well.

\section{Conclusions}
\label{sec:conc}

We conclude by recapping our main findings. 

First, we put into concrete terms the changes to the form and function of the Web. We argue that most content extraction methodologies, by their reliance on unrendered, downloaded HTML markup, do not count very large portion of the final rendered Web page. This is due to the Web's increasing reliance on external sources for content and data via JavaScript, iframes, and so on.

Second, we find that although wrapper induction techniques are prone to breakage and require frequent retraining, the heuristic/feature engineering extractors studied in this paper, which argued to not require training at all, are also quickly obsolete.

\subsection{Recommendations for future work}

We argue that the two findings presented in this paper be immediately addressed by the content extraction community, and we make the following recommendations.

\begin{enumerate}
\item Future content extraction methodologies should be performed on completely rendered Web pages, and should therefore be created as Web browser extensions or with a similar rendered-in-browser setup using a headless browser like PhantonJS, etc. This methodology will allow for all of the content to be loaded so that it may be fully extracted. A browser-based content extractor might operate similar to the popular AdBlock software, but only render content rather than simply removing blacklisted advertisers. Aside from executing JavaScript and gathering all of the external resources, a browser based content extractor would also allow for a visual-DOM representation that may improve extraction effectiveness. 

\item Future content extraction studies should examine Web pages and Web sites from different time periods to measure the overall robustness of the dataset. This is a difficult task and is perhaps contrary to the first recommendation because the external data from old Web pages may not be be easily rendered because the external may sources cease to exist. Nevertheless, it is possible to denote Web pages which have not changed via through Change Detection and Notification (CDN) systems~\cite{Chakravarthy2006} or through Last-Modified or ETag HTTP headers.

\item With the adoption semantic tags in HTML5, such as \texttt{section}, \texttt{header}, \texttt{main}, etc., as well as the creation of semantic attributes within the \url{schema.org} framework, it is important to ask whether content extraction algorithms are still needed at all. Many Web sites have mobile versions that streamline content delivery and a large number of content provides have content syndication systems or APIs that deliver pure content. It may be more important in the near future to focus attention on structured data extraction from lists and tables~\cite{Pimplikar2012,Gupta2009,Limaye2010,Cafarella2008,Cafarella2011} and integrating that data for meaningful analysis.
\end{enumerate}

Content extraction research has been an important part of the history and development of the Web, but this area of study would greatly benefit by considering these recommendations as they would lead to new approaches that are more robust and reliable.

\tg{I would somehow expect at least one last sentence to conclude the paper ... something like CE research would greatly benefit from considering these recommendations as would lead to approaches which are more robust and reliable not only across multiple websites but also across time.}

\balance

\end{document}